\documentclass{aa}
\usepackage{natbib}
\usepackage{graphicx}
\bibpunct{(}{)}{;}{a}{}{,} % to follow the A&A style

\begin{document}

\title{The absence of sub-minute periodicity in classical T Tauri stars\thanks{based on observations obtained at the Southern African Large Telescope (SALT) and Skinakas observatory, Greece}}

\author{H.~M.~G\"unther \inst{1} \and N.~Lewandowska \inst{1} \and M.~P.~G.~Hundertmark \inst{2} \and H.~Steinle \inst{3} \and J.~H.~M.~M. Schmitt \inst{1} \and D.~Buckley \inst{4} \and S.~Crawford \inst{4} \and D.~O'Donoghue \inst{4} \and P.~Vaisanen \inst{4}}
\offprints{H.~M. G\"unther,\\ \email{moritz.guenther@hs.uni-hamburg.de}}
\institute{Hamburger Sternwarte, Universit\"at Hamburg, Gojenbergsweg 112, 21029 Hamburg, Germany \and
Institut f\"ur Astrophysik, Universit\"at G\"ottingen, Friedrich-Hund-Platz 1, 37077 G\"ottingen, Germany \and
Max-Planck Institut f\"ur extraterrestrische Physik, Giessenbachstrasse 1, 85741 Garching bei M\"unchen, Germany \and
South African Astronomical Observatory, Observatory Road, Observatory 7925, South Africa
} %\email{moritz.guenther@hs.uni-hamburg.de}}
%sa@salt.ac.za
\date{Received 6 Januray 2010 / Accepted 4 May 2010}
\abstract{Classical T~Tauri stars (CTTS) are young, late-type objects, that still accrete matter from a circumstellar disk. Analytical treatments and numerical simulations predict instabilities of the accretion shock on the stellar surface.}{We search for variability on timescales below a few minutes in the CTTS TW~Hya and AA~Tau.}{TW~Hya was observed with SALTICAM on the Southern African Large Telescope (SALT) in narrow-band filters around the Balmer jump. The observations were performed in slit mode, which provides a time resolution of about 0.1~s. For AA~Tau we obtained observations with OPTIMA, a single photon-counting device with even better time resolution.}{Small-scale variability typically lasts a few seconds, however, no significant periodicity is detected. We place a 99\% confidence upper limit on the pulsed fraction of the lightcurves. The relative amplitude is below 0.001 for TW~Hya in the frequency range 0.02-3~Hz in the 340~nm filter and 0.1-3~Hz in the 380~nm filter. The corresponding value for AA~Tau is an amplitude of 0.005 for 0.02-50~Hz.}{ The relevant timescales indicate that shock instabilites should not be seen directly in our optical and UV observations, but the predicted oscialltions would induce observable variations in the reddening. We discuss how the magnetic field could stabilise the accretion shock.}
\keywords{Accretion, accretion disks -- Stars: individual: TW Hya -- Stars: individual: AA Tau -- Stars: oscillations -- Stars: variables: T Tauri, Herbig Ae/Be}

\maketitle

\section{Introduction}
\label{introduction}
Classical T~Tauri stars (CTTS) are young ($<10$~Myr), low-mass ($<3M_{\sun}$) pre-main sequence objects. They still actively accrete from a circumstellar disk. The disk does not reach down to the central star, but is truncated by the stellar magnetic field approximately at the co-rotation radius \citep{1983ards.proc..625U,1988ApJ...330..350B,1991ApJ...370L..39K}. The high-energy photons from the star irradiate, heat and ionise the inner parts of the disk, while the matter is funneled along the magnetic field lines and impacts on the star at close to free-fall velocity. For the simple case of a dipolar field, the accretion spots are located at high stellar latitude \citep{1991ApJ...370L..39K,1994ApJ...429..781S}. A strong shock develops on or near the stellar surface, which heats up the accreted matter to a few $10^6$ K. Observational support for this model comes from different wavelength bands. The disk is seen in molecular hydrogen emission lines \citep{2002ApJ...572..310H}, and the inner holes can be resolved interferometrically with radio observations \citep{2006ApJ...637L.133E}. The accretion funnel emits strongly in H$\alpha$ \citep{1998ApJ...492..743M}. This is one of the defining characteristics of CTTS. The accretion spot on the stellar surface is hot enough to emit UV and X-ray radiation \citep{lamzin, acc_model}. It also heats the underlying stellar photosphere to about $20\;000$~K \citep{calvetgullbring}.

This variety of different phenomena leads to time variability on different scales. Long changes occur over years, possibly related to a slowly changing circumstellar extinction \citep{2007A&A...461..183G}. For many CTTS the dominant source of optical variability is the stellar rotation with a typical period of a few days \citep{1983ApJ...267..191R,1986A&A...165..110B,2009ApJ...694L.153N}. Depending on its inner radius, the rotation timescale of the inner disk and also the free-fall time from the disk onto the star is a few hours. Thus variability in the accretion rate should be limited to this timescale; indeed, variations of about 0.1~mag are regularly observed in broad-band photometry from CTTS \citep{1996MNRAS.282..167S,2008MNRAS.391.1913R}. 

Nevertheless, much faster variations can be expected if the accretion shock on the stellar surface is unstable. This has first been studied in the context of accretion onto white dwarfs \citep{1983ASSL..101..199L,1982ApJ...261..543C,1985ApJ...296..128I}. Even if the magnetic field is strong enough to channel the accretion stream and prevent turbulence on the boundary between accretion stream and surrounding photosphere, the shock can still fluctuate up and down. The infalling matter builds up a funnel above the stellar surface. Owing to the ram pressure of the infalling matter, the gas gets compressed and accumulates. At some point the accumulated matter is dense enough for a runaway cooling to set in and the column collapses fast. A new shock forms at the stellar surface and works its way upwards again building up a new reservoir of matter. \citet{2008A&A...491L..17S} and \citet{2009arXiv0912.1799O} predict a typical timescale of 10~min for the fluctuations, and the same is seen in models of \citet{2008MNRAS.388..357K}, but with much shorter periods of 0.02-0.2~s. The difference can be mainly attributed to different cooling functions and different input densities $n$. Since the cooling is proportional to $n^2$, different pre-shock densities heavily influence the cooling time. \citet{2008MNRAS.388..357K} predict the shock luminosity to fluctuate about 50\% around the mean, but caution that this value and the exact shape of the fluctuation are very model dependent. They also find that magnetic fields tend to stabilise the accretion shock, so that no oscillations occur.

The most direct tracer of the accretion shock is the soft X-ray emission. This is formed just behind the shock front, where the accreted material is heated to a 2-3 MK \citep{lamzin,acc_model}. \citet{2009ApJ...703.1224D} used \emph{XMM-Newton} X-ray data to search for periodicity in \object{TW Hya}, one of the best-studied CTTS. They did not detect any period and set an upper limit on the pulsed fraction of 5\% at the 99\% confidence level over the interval 0.001-7~Hz. TW~Hya is the only target where the X-ray count rate is high enough to perform timing analyses on scales below a minute. In this paper we use a secondary tracer of the accretion shock. We obtained observations in the blue band, using optical telescopes. In the blue and the UV range, CTTS are dominated by the continuum emission from the accretion shock \citep{1974MNRAS.168..603L,1980Afz....16..243G,2000ApJ...535L..47M} because they are late-type objects and their photospheric emission quickly drops towards shorter wavelengths. It is not clear from the current models how fast the optical emission reacts to shock instabilities, but observations in the visible potentially provide a diagnostic with high signal-to-noise ratio, and yield a complementary view on sub-minute variability.

We will first introduce the targets observed in Sect.~\ref{targets} and the observations performed in Sect.~\ref{observations}. Details of the data reduction process are given in Sect.~\ref{datareduction}. In Sect.~\ref{periodicity} we search for periodicities in the data employing Lomb-Scargle periodograms. We discuss our results in Sect.~\ref{discussion} and summarise them in Sect.~\ref{summary}.

\section{Targets}
\label{targets}
\subsection{TW Hya}
TW~Hya is the closest known CTTS with a distance of only 57~pc \citep{1998MNRAS.301L..39W}. Because of the absence of a surrounding molecular cloud, TW~Hya is a key system for the study of CTTS. 
Broad H$\alpha$ profiles (FWHM $\sim 200$~km~s$^{-1}$) were observed by \citet{2000ApJ...535L..47M}. TW~Hya's  stellar parameters are $M_*\sim0.7\;M_{\sun}$, $R_*\sim1.0\;R_{\sun}$, and its age is 10~Myr \citep{1999ApJ...512L..63W}. The spectral type of TW~Hya is K7$\;$V-M1$\;$V \citep{1999ApJ...512L..63W,2002ApJ...580..343B}. It is long known that the system is seen nearly pole-on \citep{1997Sci...277...67K,2000ApJ...534L.101W,2002ApJ...571..378A,2002ApJ...566..409W} with the most precise inclination of  $7\pm1^{\circ}$ measured by \citet{2004ApJ...616L..11Q}. Moreover, TW~Hya displays variations in line profiles and veiling, which have been interpreted as signatures of accretion spot rotation \citep{1988AJ.....96.1949H,1994AJ....108.1906H,1998A&A...340..135M,2002ApJ...571..378A,2002ApJ...580..343B}.

Although relatively old for a CTTS ($\approx10$~Myr), it is still actively accreting from a surrounding disk. 
Modelling of the spectral energy distribution from TW~Hya requires a gap in the inner disk \citep{2002ApJ...568.1008C} and \citet{2006ApJ...637L.133E} interferometrically resolve its disk and confirm the size of the inner hole. \citet{2002ApJ...572..310H} observed TW~Hya in the UV and \citet{2004ApJ...607..369H} constrain the inner disk temperature to about 2500~K using H$_2$ fluorescence in \emph{HST/STIS} and \emph{FUSE} data.
Early photometric observations show variability between magnitude 10.9 and 11.3 in the V-band \citep{1983A&A...121..217R}. The most detailed photometric study so far found variability on timescales of hours to days with flicker noise probabilities, where the amplitude of variations $a$ depends on the frequency $f$ as $a\propto 1/\sqrt{f}$ for periods between ten days and a few hours ($10^{-6}-10^{-4}$~Hz) \citep{2008MNRAS.391.1913R}. TW~Hya also was an early target for X-ray observations with \emph{ROSAT} \citep{2000A&A...354..621C}, \emph{Chandra} \citep{2002ApJ...567..434K,2009A&A...505..755R,2010ApJ...710.1835B} and \emph{XMM-Newton} \citep{twhya}. In the X-ray range no period was found and 
\citet{2009ApJ...703.1224D} placed a 99\% confidence limit that, at most, 5\% of the signal is pulsed in the frequency range 0.0001-6.81~Hz.

\subsection{AA Tau}
\object{AA Tau} is located in the Taurus-Auriga star forming region in a distance of about 140~pc. According to its parameters it is a typical CTTS with spectral type K7 \citep{1979ApJS...41..743C}, a mass of $\sim0.8\;M_{\odot}$, and radius $\sim1.85\;R_{\odot}$. In contrast to TW~Hya, AA~Tau is observed nearly edge-on \citep{2005MNRAS.358..632O} and periodically a warp in the disk occults the star, causing brightness variations of 1.4~mag \citep{1977ApJ...214..747H,1989AJ.....97..483V,1993AJ....106.1608V,1999A&A...349..619B,2003A&A...409..163M}. The period is 8.22~days, which seems to be the stellar rotation period \citep{2007A&A...463.1017B}. The occultation is not strictly periodic, but also variable over longer timescales. In some instances the fading of the star did not occur; this hints at changing accretion geometries \citep{2003A&A...409..169B}. The occultation of the UV and optical emission coincides with increased column densities seen in X-ray observations \citep{2007A&A...462L..41S,2007A&A...475..607G}. Apart from one bright, coronal flare the X-ray luminosity is not sufficient to repeat the analysis done for TW~Hya.

\section{Observations}
\label{observations}
\subsection{TW Hya}
TW~Hya was observed with the Southern African Large Telescope (SALT) for 45~min in total starting at UTC~23:59 on 28-03-2009. The spherical SALT primary mirror with a total diameter of about 11~m is segmented, fixed in elevation, and tilted about 37\degr{} from the vertical. This means that the telescope can observe only an annulus of the sky, which is 12 degrees wide. The CCD camera, SALTICAM, is positioned in a prime focus ``payload'', which moves across a virtual spherical focal surface during an observation, which can last up to $\approx$2 hours. The full observation of TW~Hya was scheduled in one crossing. To achieve the highest possible time resolution we used the slot mode of SALTICAM, where most of the CCD is covered by a mask and only a small strip of 144~rows is read out. With an on-chip binning of $6\times 6$, an effective sampling period of about 0.14~s is reached. SALTICAM consists of two CCD chips with two amplifiers each. The CCDs are separated by about 1.5~mm and each of the amplifiers has a different bias level. Observations were taken in three different bands for 15~min each. The filters used, in order of observation, are U-band and two narrowband UV interference filters with central wavelength 340~nm (FWHM 35~nm) and 380~nm (FWHM 40~nm). TW~Hya is severely overexposed in the U-band, thus this data was not included in the analysis. At the time of the observations SALTICAM operated with software version 4.78.
More details on SALT and the instrument SALTICAM  are presented in \citet{2008SPIE.7014E...6B}.

The design of SALT has some implications for photometry: Only differential photometry is possible since the entrance pupil of the telescopes moves with respect to the primary mirror array, thus continuously changing the effective collecting area.
Also, the light is partially vignetted in the slot mode due to the relative position of the slit mask and the CCD. Thus the intensity varies as a function of the CCD row. 

\subsection{AA Tau}
AA~Tau was observed with the high speed photo-polarimeter OPTIMA Burst (Optical Pulsar TIMing Analyzer) on the 1.3~m Cassegrain telescope of the Skinakas Observatory (SKO) in Greece on 07-10-2008 from 02:15 to 03:27~UT.  
OPTIMA Burst has several possible configurations.
The photometer consists of a single photon counting fibre-fed system that contains six fibre apertures which are placed in a hexagonal form around the central fiber for the target.
The apertures have a diameter of 300 micrometers each. This corresponds to 6 arcsec on the sky with the SKO telescope.
An additional fibre is placed about 1 arcminute away from the fibre bundle and is used to determine the sky background.
OPTIMA Burst has an intrinsic photon-arrival-time resolution of 4 microseconds and can record white light radiation in the wavelength range from 450 to 900~nm. AA Tau was observed with a blue filter (350-550~nm). The fibres are placed in a slanted mirror in the prime focus of the telescope. The incoming light is also partly reflected by the slanted mirror to a fast-readout Apogee AP6 CCD camera with a Kodak chip of KAF1000E type.
It continuously takes exposures with exposure time 10 seconds while the telescope is tracking. They are used to position the target in the central fiber of the hexagonal bundle, thus controlling the tracking, and to monitor the seeing conditions during the observations.
The arrival times of the photons in the fibre system are correlated with a GPS timing receiver. A more detailed description of the technical properties of OPTIMA Burst is given by \citet{2008ASSL..351..153K}.

\section{Data reduction}
\label{datareduction}
\subsection{TW Hya}
\label{dataredtw}
SALTICAM data reduction was performed with the PySALT package\footnote{\texttt{http://www.salt.ac.za/science-support/}}, which is provided by the SALT Astronomy Operations team and runs in the PYRAF environment, a Python extension to the well-known IRAF system. Due to a bug in the SALTICAM software up to version 4.78 the timing information of the exposures was inaccurate. This was corrected using the task \texttt{slotutcfix}. We merged the data read out from all four amplifiers into single frames with \texttt{slotmerge}, because the best available comparison star is recorded on a different amplifier than our target. We extracted the target and comparison star using a simple square aperture with side length of 10 pixels in each coordinate to cover the entire point-spread function (PSF). The background was median filtered and estimated from an annular box centred around the target with inner and outer dimensions of 16 and 24 pixels, respectively. The fluxes of TW~Hya and one (340~nm filter) or three comparison stars (380~nm filter) were extracted with the task \texttt{slotphot}, which centres the extraction region on the target for each exposure. In a few exposures the algorithm cannot identify one of the stars and extracts a background region only. Those frames can easily be identified because the flux drops dramatically in the respective lightcurve for a single exposure. Those frames are removed from the lightcurve.

TW~Hya is a factor of 10 and 6.5 brighter than the brightest available comparison star in the filters centred on 340~nm and 380~nm, respectively. Thus, the errors on the relative photometry are dominated by the noise in the lightcurve of the comparison star. To detect periodic signals we use the higher signal-to-noise lightcurve of TW~Hya directly and search for periods in the flux ratio of both stars independently. The occurrence of the same period in more than one star would be a strong indicator for an instrumental effect.

\begin{figure}
\resizebox{\hsize}{!}{\includegraphics{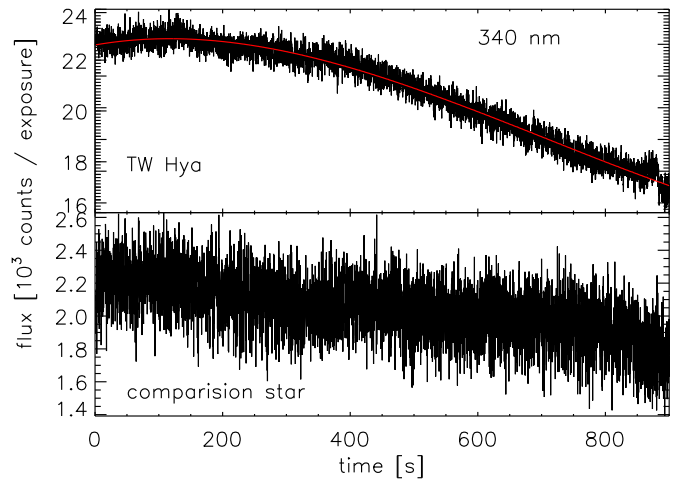}}
\caption{Flux in the filter centred on 340~nm for TW~Hya and the comparison star. The red/grey line shows a polynomial fit.}
\label{TW_Hya_lc_34}
\end{figure}
\begin{figure}
\resizebox{\hsize}{!}{\includegraphics{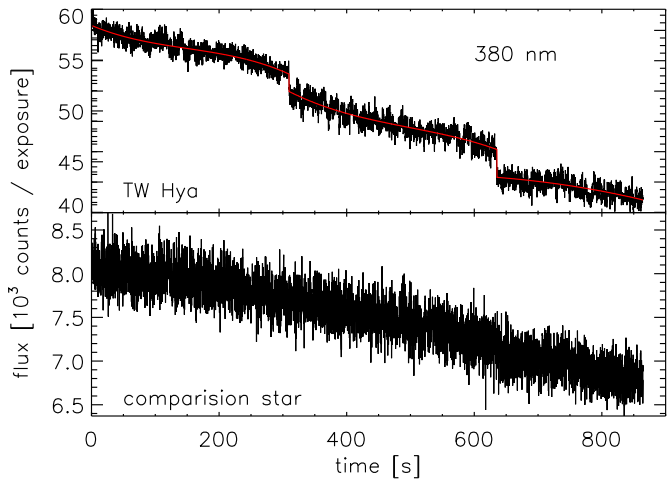}}
\caption{Flux in the filter centred on 380~nm for TW~Hya and a brightest of three comparison stars.  The red/grey line shows a polynomial fit (see text for more details).}
\label{TW_Hya_lc_38}
\end{figure}
We show the lightcurves of TW~Hya and one comparison star for both filters in Figs.~\ref{TW_Hya_lc_34} and \ref{TW_Hya_lc_38}. The lightcurve of TW~Hya in Fig.~\ref{TW_Hya_lc_38} contains two steps after about 300~s and 650~s, which are not easily visible in the lightcurve of the comparison star. The later step is likely present there, too, but hidden in the higher noise level. These steps are due to repositioning the target in the slot (an autoguider was not yet available). Such a repositioning did not take place during the observation in the 340~nm filter. TW~Hya and the comparison star are affected differently, because they have different coordinates on the CCD chip, thus the steps are still present in the relative lightcurve. This introduces spurious signals for long periods and we therefore limit our analysis to the frequency range 0.02-3~Hz in the 340~nm band and 0.1-3~Hz in the 380~nm band. We fitted the lightcurve of TW~Hya with a third order polynomial over the entire 340~nm exposure (red/grey line in Fig. \ref{TW_Hya_lc_34}) and with separate third order polynomials for each step in the 380~nm exposure (red/grey line in Fig. \ref{TW_Hya_lc_38}) to remove trends before further analysis. 

\subsection{AA Tau}
\begin{figure}
\resizebox{\hsize}{!}{\includegraphics{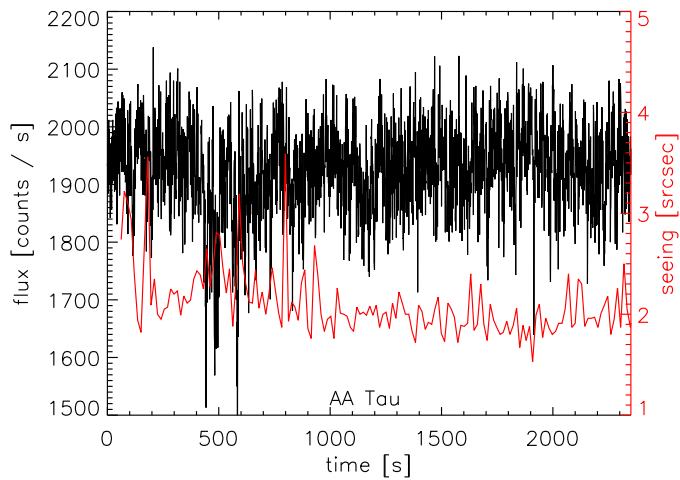}}
\caption{Count rate for AA Tau, binned to 1~s (black - left axis) and FWHM of a comparison star in the CCD image with 10~s time resolution (red/grey - right axis).}
\label{AATau_lc}
\end{figure}

We use the data from the central fiber only, which contains essentially all the signal. The lightcurve shows some drop-outs, where no data was read out and a strong increase towards the end of the night. We selected a clean, continuous region of about 40~min before the twilight sets in for analysis. This part of the lightcurve is shown in Fig.~\ref{AATau_lc}, binned to 1~s.
The second line shows the evolution of the full width at half maximum (FWHM) as fitted to a comparison star in the accompanying CCD images (timing resolution 10~s). Times with large FWHM always correspond to low flux because light is lost from the instrument. We cannot correct for this effect by means of a comparison star, because the CCD requires an exposure time of 10~s, which is much longer than the sampling rate of the target. The lightcurve is mean subtracted before further analysis.

\section{Analysis}
\label{periodicity}
We determine the typical timescale of fluctuations by the autocorrelation function.
To search for periodicity we first calculated Lomb-Scargle periodograms of our lightcurves. Additionally we performed a wavelet analysis, because periodic signals with changing frequency would be washed out in a simple periodogram. We quantify the significance of our non-detection by analytic considerations and Monte-Carlo simulations.
\subsection{Autocorrelation function}
\begin{figure}
\resizebox{\hsize}{!}{\includegraphics{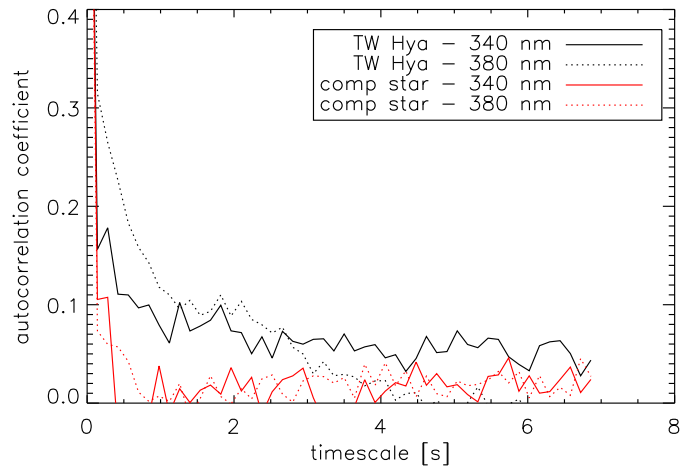}}
\caption{Autocorrelation coefficient for TW~Hya and the brightest comparison star in the field in both bands.}
\label{TW_Hya_autocorr}
\end{figure}

The autocorrelation function is defined as
\begin{equation}
P(L)=\frac{\sum_{j=1}^{j=N-L}X_j X_{j+L}}{\sum_{j=1}^{j=N} X_j^2},
\end{equation}
where $X_j$ are the values of the de-trended and mean-subtracted lightcurve and $N$ is the total number of points. $L$ is the scale of the correlation. A correlation coefficient of 1 indicates a perfect correlation, 0 the absence of a linear correlation and -1 a perfect anti-correlation. We show the auto-correlation coefficients for TW~Hya in Fig.~\ref{TW_Hya_autocorr} for both filters. 
% \begin{figure}
% \resizebox{\hsize}{!}{\includegraphics{AATau_autocorr}}
% \caption{Autocorrelation coefficient for AA~Tau and the background during stable atmospheric conditions.}
% \label{AATau_autocorr}
% \end{figure}
The autocorrelation of scale 0 is always 1; in TW~Hya the coefficient differs significantly from the coefficient calculated for a comparison star for 2-4~s. The autocorrelation function for AA~Tau is different from zero for about 10~s, however we find similar results using an intrinsically stable white dwarf from a different observation as comparison object. Apparently 10~s is the timescale of atmospheric variability at the SKO for certain weather conditions. The lightcurves of the comparison stars for AA~Tau are not sufficiently densly sampled to test for this.

\subsection{Lomb-Scargle periodograms}
We calculated normalised Lomb-Scargle periodograms from our lightcurves with an IDL code\footnote{available at\\ \texttt{http://astro.uni-tuebingen.de/software/idl/aitlib/}}, which follows the method described in \citet{1982ApJ...263..835S} and \citet{1986ApJ...302..757H}. The power $P$ at a given angular frequency $\omega$ is calculated as
\begin{eqnarray}
P(\omega)=&\frac{1}{2}\frac{\left[ \sum_j X_j \cos \omega(t_j-\tau)  \right]^2}{\sum_j \cos^2 \omega(t_j-\tau)}\\
& +\frac{1}{2} \frac{\left[ \sum_j X_j \sin \omega(t_j-\tau)  \right]^2}{\sum_j \sin^2 \omega(t_j-\tau)}, \nonumber
\end{eqnarray}
where $j$ runs over all $N$ flux values $X_j$ observed at time $t_j$. $\tau$ is defined by
\begin{equation}
\tan(2 \omega \tau) = \left(\sum_j \sin 2 \omega t_j \right) / \left(\sum_j \cos 2 \omega t_j \right)
\end{equation}
In order to normalise the periodograms they are scaled with the total observed variance \citep{1986ApJ...302..757H}.
If the measurement errors are Gaussian distributed, the statistics allow us to calculate a false alarm probability (FAP) for each power level. The FAP is the probability that a lightcurve of pure noise exhibits at least one frequency with the specified power. The value \mbox{1-FAP} can thus be treated as the significance of the detection of any periodic signal. 

For the simple case of a sinusoidal signal with amplitude $A$ and angular frequency $\omega_0$, which is hidden in noise with variance $\sigma_N^2$, the normalised power $P_N(\omega_0)$ is given by \citep{1986ApJ...302..757H}:
\begin{equation}
P_N(\omega_0)=\frac{N_0}{2}\left(1+2\frac{\sigma_N^2}{A^2}\right)^{-1} \; . \label{eqn_pn}
\end{equation}

\subsubsection{TW Hya}
The sampling rate of SALTICAM is 0.14~s, thus the Nyquist frequency is just above 3~Hz.
\begin{figure}
\resizebox{\hsize}{!}{\includegraphics{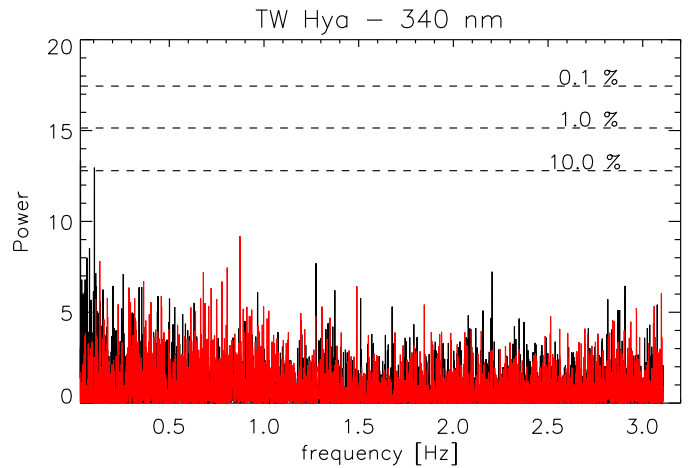}}
\caption{Lomb-Scargle periodogram of the lightcurve of TW~Hya (black) and the relative lightcurve of target and comparison star (red/grey) taken in the 340~nm filter. Dashed lines indicate false alarm probabilities.}
\label{TW_Hya_period_34}
\end{figure}
\begin{figure}
\resizebox{\hsize}{!}{\includegraphics{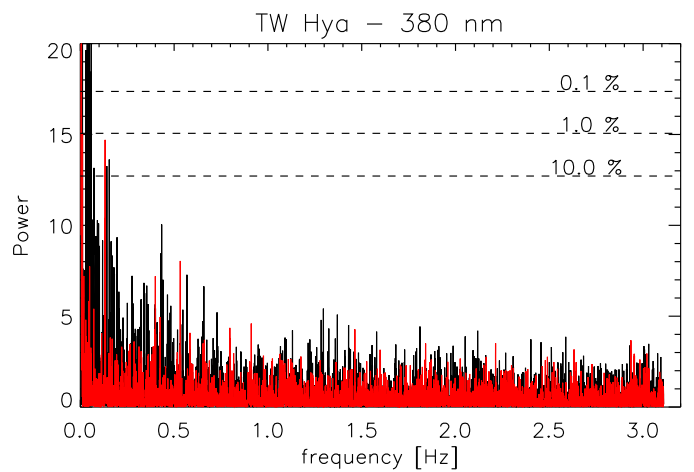}}
\caption{Lomb-Scargle periodogram of the lightcurve of TW~Hya (black) and the relative lightcurve of target and comparison star (red/grey) taken in the 380~nm filter. Dashed lines indicate false alarm probabilities.}
\label{TW_Hya_period_38}
\end{figure}
In Figs.~\ref{TW_Hya_period_34} and \ref{TW_Hya_period_38} the periodogram of the lightcurves of TW~Hya and the relative lightcurve TW~Hya and comparison star are shown. The large peaks at low frequencies in the second figure are due to the steps in the lightcurve, which we could not fully correct. No period is detected at the 99\% confidence level. There are less significant peaks around 0.135~Hz in the periodogram of the relative lightcurve and 0.14~Hz and 0.155~Hz in the absolute lightcurve in the 380~nm filter. The expected uncertainty in frequency for the given length of the observation is of the order of mHz \citep[our eqn.~\ref{eqn_pn} and][]{1981Ap&SS..78..175K}, so these frequencies differ significantly, where real signals should agree; they could possibly be due to aliasing from the large low-frequency peaks.
\begin{figure}
\resizebox{\hsize}{!}{\includegraphics{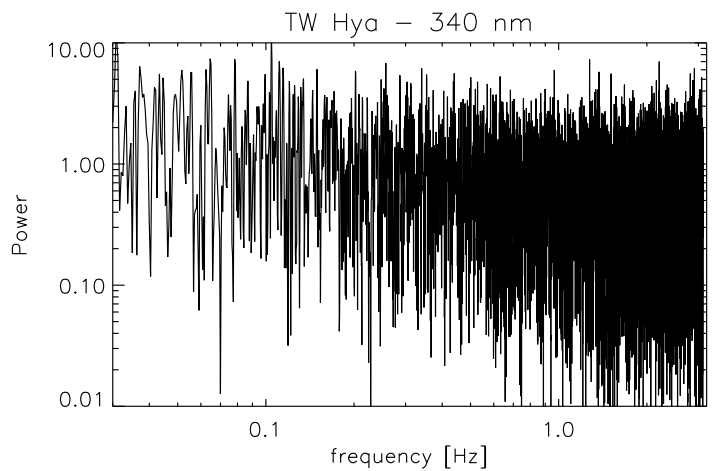}}
\caption{Same as Fig.~\ref{TW_Hya_period_34} on logarithmic scales}
\label{TW_Hya_period_34loglog}
\end{figure}
In Fig.~\ref{TW_Hya_period_34loglog} one periodogram is plotted on logarithmic scales. In contrast to \citet[][see their Figs. 3 and 7]{2008MNRAS.391.1913R}, who probe periods from hours to weeks, we do not observe a flicker noise but find the power level to be independent of the frequency.

\subsubsection{AA Tau}
\begin{figure}
\resizebox{\hsize}{!}{\includegraphics{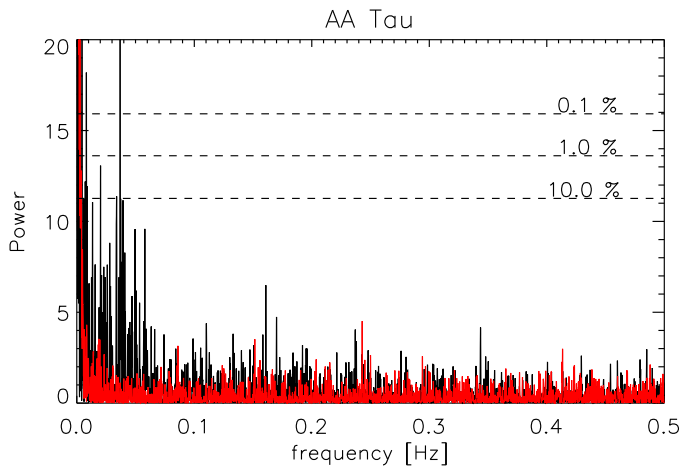}}
\caption{Lomb-Scargle periodogram of the lightcurve of AA~Tau (black) and the background lightcurve (red/grey).}
\label{AATau_period}
\end{figure}

\begin{figure}
\resizebox{\hsize}{!}{\includegraphics{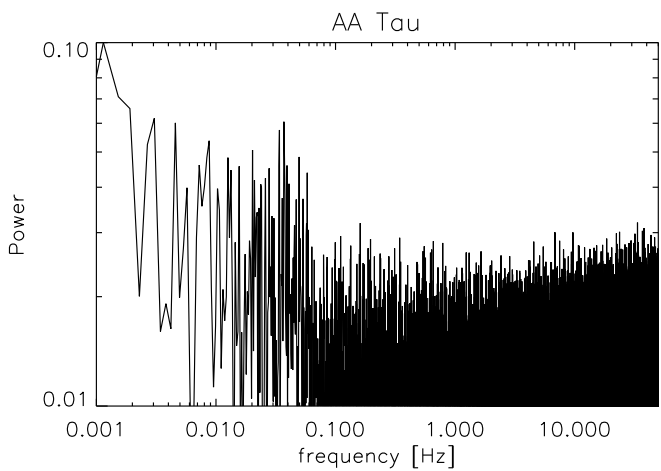}}
\caption{Fourier transform of the lightcurve of AA~Tau sampled at 100~Hz.}
\label{AATau_periodloglog}
\end{figure}

Figure~\ref{AATau_period} shows a Lomb-Scargle periodogram of the AA~Tau lightcurve, which was binned to 1~s, and of the corresponding background data. AA~Tau shows significant peaks at very low frequencies, which are also seen in the background. There is another, highly significant, peak at 0.038~Hz; however, this is a known instrumental signal probably caused by feedback between the autoguider and the telescope positioning. We conclude that there is no significant periodicity between 0.05 and 0.5~Hz.
A Lomb-Scargle periodogram of the full dataset is computationally too expensive, instead we calculated the Fast Fourier transform (FFT) of the full lightcurve sampled at 100~Hz (Fig.~\ref{AATau_periodloglog}).
The largest peak seen is again the systematic signal at 0.038 Hz (as already discussed) and we can exclude significant periodicity down to 50~Hz, which is the Nyquist frequency for a sampling rate of 100~Hz.

\subsection{Wavelet analysis}
If periodic oscillations exist in the accretion shocks of CTTS their frequency may be unstable. Therefore we performed a wavelet analysis of the lightcurves using the IDL function \texttt{wv\_applet}. Evenly sampled lightcurves are required as  input, thus missing data points were filled with repeated values of the last valid flux measurement. We employed different wavelet shapes, concentrating on Morlets of order 4-6. For TW~Hya no signal with a significance above 90\% is found in the wavelet analysis, while AA~Tau shows a peak in the power spectrum after about 500~s, which is very likely caused by the dip in the lightcurve (Fig.~\ref{AATau_lc}). We showed above that this feature is related to changing seeing conditions.

\subsection{Upper limits on periodicity}
In TW~Hya we do not detect periodic signals on the 99\% confidence level, that is power level $P_N= 17.4$. Knowing the variance of the noise $\sigma_N^2$ we now invert eqn.~\ref{eqn_pn} to calculate the minimum amplitude $A_m$ which should have been detected at that level, if present:
\begin{equation}
A_m=\sigma_N \left(\frac{N_0}{4 P_N}-\frac{1}{2}\right)^{-\frac{1}{2}} \label{snr}
\end{equation}
For the parameters of both lightcurves from TW~Hya $A_m\approx10 \sigma_N$. The total noise level in the normalised lightcurves is about $10^{-2}$ and $5\times 10^{-3}$ in the 340~nm and 380~nm observations of TW~Hya, therefore the 99\% confidence limit on $A_m$ is $10^{-3}$ and $5\times 10^{-4}$, respectively for the presence of a single, sinusoidal mode in the frequency range 0.02-3~Hz in the 340~nm band and 0.1-3~Hz in the 380~nm band.

We performed Monte-Carlo simulations to estimate the detection efficiency for a combination of many periods. We used the time bins from the observations, keeping the irregularities introduced by the outlier rejection and simulated 1000 lightcurves, calculated Lomb-Scargle periodograms for each of them, and determined the largest peak in the power spectrum. For random measurement uncertainties of 1\%, which corresponds to the observations of TW~Hya in the 340~nm filter, we distributed multiple sinusoidal waves with random frequencies between 0.02 and 3~Hz and added them to a constant component. In the case of 40 different frequencies, each of them with an amplitude of 5\% of the constant component, more than 90\% of the simulations produced at least one peak in the power spectrum above the 10\% FAP, that is higher than in our observations (85\% of the simulations exceeded 99\% FAP). 
For smaller amplitudes of only 2\% of the continuum level the peaks in the power spectrum are smaller, still 92\% of the simulations with 10 different frequencies produced at least one signal above the 10\% FAP (91\% of the simulations exceeded 99\% FAP). Due to the larger signal and consequently lower uncertainties these values are even better for the data taken with the 380~nm filter.

Due to the lower signal-to-noise ratio, the data for AA~Tau is not quite as sensitive. Equation~\ref{snr} shows, that a signal of 0.13 of the measurement error, that is about $5\times10^{-3}$ of the total flux, should have been detected at 99\% confidence level in AA~Tau. Since the noise level is about three times larger than in TW~Hya correspondingly larger amplitudes are needed to detect multiple frequencies.

\section{Discussion}
\label{discussion}
Our analysis of the data presented above yields no significant periodicity in the searched frequency range. The limits set by this data on the maximum amplitude for variations are up to two orders of magnitude more sensitive than those \citet{2009ApJ...703.1224D} derived for the X-ray lightcurve of TW~Hya. However, the soft X-rays are formed directly in the shock cooling zone with a weak contribution from the stellar corona \citep{acc_model}, therefore much stronger oscillations are expected in X-rays \citep{2008A&A...491L..17S} than in the UV wavelength range we probe. So far no time-dependent radiative-transfer calculations of accretion shocks in CTTS have been carried out. In stationary 1-D models the hot gas at the lower end of the accretion column and the heated photosphere below  emit a significant portion of their energy around the Balmer-jump \citep{calvetgullbring}. In this wavelength region the accretion spot dominates  over the stellar photosphere by one order of magnitude in TW~Hya \citep{2000ApJ...535L..47M}. The heated photosphere is optically thick and can be treated as a black-body with temperatures between 5000~K and 8000~K \citep[see Fig.~3 in][]{calvetgullbring}, its luminosity should thus be of the order of $10^{11}$~erg~s$^{-1}$~cm$^{-2}$.  The \citet{calvetgullbring} model refers to younger CTTS with a radius of $2 R_{\sun}$, however the spectral type of TW~Hya matches the model, thus we can apply it to obtain an order-of-magnitude estimate. The model implies a column mass of the heated photosphere down to the region, where the continuum is formed, of the order 10~g~cm$^{-2}$, containing a thermal energy of $10^{13}$~erg~s$^{-1}$~cm$^{-2}$, and suggests a timescale for cooling of the heated photosphere of 100~s. This calculation gives only a rough estimate, because the cooling would change the temperature and thus the opacity structure, so the deeper layers become visible as the cooling proceeds. It suggests that rapid changes of the shock conditions will not be visible in optical and UV emission from the heated photosphere; however, detailed radiative transfer calculations are required to confirm this.

According to the models the flow time of accreted matter through the post-shock zone is a few seconds. Incidentally, the flow time through the shock cooling zone matches the time we find in the autocorrelation function for TW~Hya.  
The short term variability in the lightcurves shown in figures~\ref{TW_Hya_lc_34}-\ref{AATau_lc} is dominated by statistical fluctuations, so the source variability, which is detected in the autocorrelation function, is of the order of 1\% of the continuum level at most. Several effects could be responsible for that: If the accretion flow is inhomogeneous and clumpy, this would cause the UV emission from the shock cooling zone to vary within seconds. The size of the accreted clumps is presumably small in comparison to the total area covered by accretion spots, which is estimated from 0.3\% to 5-10\% of the stellar surface for TW~Hya, depending on the wavelength of the observation \citep[][and references therein]{acc_model}. If the clumps switch on or off the emission from 1\% of the accretion spot, then their characteristic size is $10^3-10^4$~km. The sound speed in the hot parts of the post-shock cooling zone is of the order 100~km~s$^{-1}$, thus an area with radius $10^3$~km could be affected within a few seconds; in cooler regions of the post-shock cooling zone, where most of the optical emission originates, the temperature and thus the sound speed are about an order of magnitude lower. In this region only very small areas can respond coherently to disturbances.

Instead of inhomogeneous accretion the sound waves could also be caused by accretion shock instabilities. Three-dimensional simulations are needed to study this in more detail. But would the optical emission react fast enough to changing conditions, however they are caused? Even if it does not, the oscillations of the accretion shock could be seen in absorption.
In the time-dependent simulations of \citet{2008A&A...491L..17S} the accretion shock oscillates up and down. Thus, the pre-shock column density changes with time by $\approx10^{20}$~cm$^{-2}$. This accreted material is not reprocessed in the hot shock yet and we apply the standard conversion between X-ray column density and optical absorption $A_V= N_H/ 2\times 10^{21}$~cm$^{-2}=0.05$~mag \citep{2003A&A...408..581V}. According to \citet{1999PASP..111...63F} an additional absorption of $A_V=0.05$~mag would reduce the flux at 3800~\AA{} by 7\%, much more than our upper limit on periodic variability. The numbers given are calculated for a line-of-sight along the accretion flow. TW~Hya is seen pole-on, so this condition is valid for accretion at high latitude. The extinction changes with the viewing geometry; the argument does not apply if the accretion column is seen from the side.

While general variability is seen, clear periods are absent with significant limits down to 0.001 for the maximum amplitude of a pulsed fraction of the emission. It might simply be hard to observe oscillations in the optical, as discussed above, but \citet{2009ApJ...703.1224D} also do not see a periodicity in the X-ray lightcurve of TW~Hya. Those authors present a detailed discussion on the expected period of the oscillations, the effect of multiple accretion streams and the influence of the magnetic field, arguing the supporting pressure of the magnetic field has to be taken into account.

Different studies of time-dependent accretion shocks concentrate on different aspects. \citet{2008MNRAS.388..357K} present 1D magneto-hydrodynamic models of super-Alfv\'enic accretion flows with infall velocities around 100~km~s$^{-1}$ and different incident angles. In their simulations strong magnetic fields lead to a stable accretion flow. They develop a stability criterion which depends on the angle between the field and the stellar surface. However, this criterion has not been verified for higher infall velocities. If it is still valid for velocities of 525~km~s$^{-1}$ and densities of $10^{-12}$~g~cm$^{-3}$, as measured from X-ray spectroscopy in TW~Hya \citep{2002ApJ...567..434K,twhya,acc_model,2009A&A...505..755R,2010ApJ...710.1835B}, then a magnetic field of 100~G is sufficient to keep the shocks stable for an accretion flow perpendicular to the stellar surface.  Yet, it is not clear if this is also valid for the stronger magnetic fields usually measured in the accretion spot regions of CTTS \citep{2004Ap&SS.292..619V, 2008MNRAS.386.1234D},  which indicate sub-Alfv\'enic flows. \citet{2010A&A...510A..71O} performed 2D simulations of accretion perpendicular to the stellar surface including heat conduction  in both the super-Alfv\'enic and the sub-Alfv\'enic regime. They confirm the importance of the magnetic field, and find that shocks with high $\beta$ (the ratio between gas pressure and magnetic pressure) become stable through magnetic damping  for super-Alfv\'enic flows. \citet{2010A&A...510A..71O} observe quasi-periodic shock oscillations for cases with larger magnetic fields  when the flow becomes sub-Alfv\'enic and chaotic behaviour without periodicity for even weaker magnetic fields. \citet{2010A&A...510A..71O} also compare their results to 1D simulations with a similar setup by \citet{2008A&A...491L..17S} and find that 2D models predict a smaller amplitude and higher frequency than 1D models. For large values of $\beta$ the magnetic field cannot confine the accretion stream to a funnel and matter flows out side-ways, disturbing the surrounding stellar atmosphere.

Although \citet{2008MNRAS.388..357K} and \citet{2010A&A...510A..71O} predict stability in different regimes, they both agree that there are magnetic field configurations that inhibit shock oscillations in the super-Alfv\'enic regime. Only \citet{2010A&A...510A..71O} probe sub-Alfv\'enic flows, but they test less parameters than \citet{2008MNRAS.388..357K}, so we cannot exclude that similar stability regions exit for sub-Alfv\'enic flows. So,  we may not see periodic oscillations of the accretion shock, not because of confusion of a larger number of accretion streams, but because the shock is intrinsically stable.

\section{Summary}
\label{summary}
We presented observations of TW~Hya with SALT in two filters centred just above and below the Balmer jump with a timing resolution of 0.14~s and observations of AA~Tau with the single-photon counting device OPTIMA. TW~Hya shows an autocorrelation on timescales of a few seconds, however we detect no periodic signal in TW~Hya or AA~Tau. We place a 99\% confidence upper limit on the pulsed fraction of the lightcurves as 0.001 for TW~Hya in the frequency range 0.02-3~Hz in the 340~nm filter and 0.1-3~Hz in the 380~nm filter. The corresponding value for AA~Tau is an amplitude of 0.005 for 0.02-50~Hz. We simulate lightcurves to calculate upper limits for scenarios with multiple independent accretion shocks with different periods. The timescale seen in the autocorrelation function is consistent with the flow time through the accretion shock and the dimension of affected area as estimated from the amplitudes in the lightcurve. Our limits on periodicity are so stringent, that we discuss the magnetic field needed to stabilise the accretion shock.

\begin{acknowledgements}
Half of the observations reported in this paper were obtained with the Southern African Large Telescope (SALT). The other half of the observations were obtained using OPTIMA (Optical Pulsar TIMing Analyzer), which is build and operated by the MPE, Garching, Munich and installed at Skinakas observatory at the 1.3m telescope. Skinakas Observatory is a collaborative project of the University of Crete, the Foundation for Research and
Technology - Hellas, and the Max-Planck-Institute for Extraterrestrial Physics. 

H.M.G. acknowledges support from DLR under 50OR0105 and H.M.G. and M.P.G.H. support from the DFG-funded GrK~1351.
\end{acknowledgements}

\bibliographystyle{aa} % style aa.bst
\bibliography{../articles}
\end{document}